# Comments on Ionization Cooling Channel Characteristics

David Neuffer[a]

[a]*Fermilab, PO Box 500, Batavia IL 60510*

**Abstract.** Ionization cooling channels with a wide variety of characteristics and cooling properties are being developed. These channels can produce cooling performances that are largely consistent with the ionization cooling theory developed previously. In this paper we review ionization cooling theory, discuss its application to presently developing cooling channels, and discuss criteria for optimizing cooling.

## INTRODUCTION

The MAP program (Muon Accelerator Project) is developing a number of ionization cooling channels for possible use for muon colliders. These cooling channels should be evaluated for cooling efficiency and be compared with expected cooling performance. Any deviations from expected performance should be evaluated and the causes of deviations should be determined. The general framework for the discussion of muon cooling (ionization cooling) was developed 30 years ago, where the key ingredient was the development and evaluation of damping partition numbers. Ionization cooling channels are naturally antidamping longitudinally and must be modified to enable 6-D cooling by changing the partition numbers. Since then, solenoidal and Li lens focusing have been introduced to obtain strong focusing at the damping absorbers, and helical cooling channels have been introduced for strong focusing with longitudinal damping. These developments have reinforced the importance of the fundamental ionization cooling equations, and well-designed cooling channels accurately follow their predictions. Deviations are usually associated with poor dynamic acceptance, or chromatic effects, or poor matching into a cooling section, or inadequate longitudinal (rf) focusing. It is also relatively difficult to obtain good longitudinal cooling; the longitudinal partition number must be modified to a clearly damping value and strong rf focusing is needed.

In this paper we reintroduce the ionization cooling equations, with an emphasis on the partition numbers, and with a discussion of longitudinal and transverse focusing. We apply these equations to sample cooling channels, and indicate criteria for good or better cooling. Criteria for improved and efficient cooling are discussed.

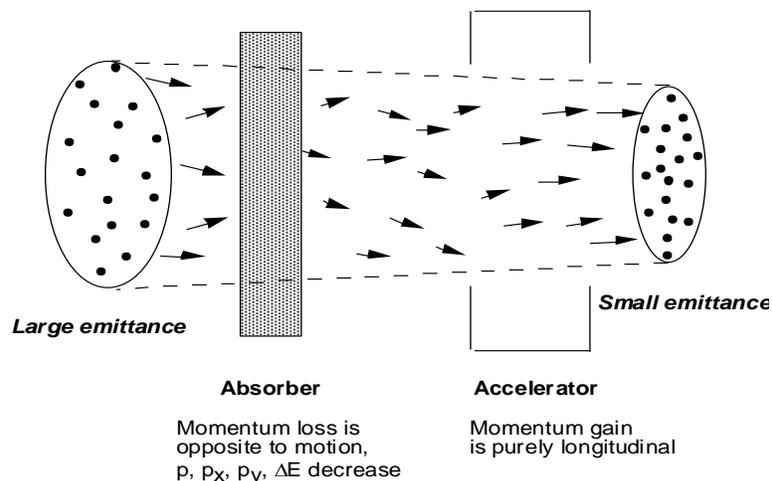

**FIGURE 1.** Concept of ionization cooling.

# MUON COOLING EQUATIONS

In ionization cooling (μ-cooling), particles pass through a material medium and lose energy (momentum) through ionization interactions, and this is followed by beam reacceleration in rf cavities.(Figure 1) The losses are parallel to the particle motion, and therefore include transverse and longitudinal momentum losses; the reacceleration restores only longitudinal momentum. The loss of transverse momentum reduces particle emittances, cooling the beam. However, the random process of multiple scattering in the material medium increases the rms beam divergence, adding a heating term which must be controlled in a complete cooling system.

The differential equation for rms transverse cooling is [1, 2, 3, 4]:

$$\frac{d\varepsilon_N}{ds} = -\frac{g_t}{\beta^2 E}\frac{dE}{ds}\varepsilon_N + \frac{\beta\gamma\,\beta_\perp}{2}\frac{d\langle\theta_{rms}^2\rangle}{ds} \quad , \tag{1}$$

where the first term is the energy-loss cooling effect and the second is the multiple-scattering heating term. Here $\varepsilon_N$ is the normalized rms emittance, $E$ is the beam energy, $\beta = v/c$ and $\gamma$ are the usual kinematic factors, $dE/ds$ is the energy loss rate, and $\theta_{rms}$ is the rms multiple scattering angle. The multiple scattering can be approximated by:

$$\frac{d\langle\theta_{rms}^2\rangle}{ds} = \frac{E_s^2}{\beta^4\gamma^2 L_R (m_\mu c^2)^2} \quad , \tag{2}$$

where $L_R$ is the material radiation length, $\beta_\perp$ is the betatron function at the absorber, and $E_s$ is the characteristic scattering energy (~13.6 MeV).[6] (The normalized emittance is related to the geometric emittance $\varepsilon_\perp$ by $\varepsilon_N = \varepsilon_\perp(\beta\gamma)$, and the beam size is given by $\sigma_x = (\varepsilon_\perp \beta_\perp)^{1/2}$.) $g_t$ is the transverse cooling partition number; $g_t = 1$ without the transverse-longitudinal coupling discussed below.

Longitudinal cooling depends on having the energy loss mechanism such that higher-energy muons lose more energy. The equation is:

$$\frac{d\varepsilon_L}{ds} = -\frac{\partial\left(\frac{dE_\mu}{ds}\right)}{\partial E_\mu}\varepsilon_L + \frac{\beta\gamma\,\beta_L}{2}\frac{d\langle(\delta p/p)^2_{rms}\rangle}{ds} = -\frac{g_L}{\beta^2 E_\mu}\frac{dE_\mu}{ds}\varepsilon_L + \frac{\beta\gamma\,\beta_L}{2}\frac{d\langle(\delta p/p)^2_{rms}\rangle}{ds} \tag{3}$$

where $\beta_L$ is a longitudinal betatron function, and we are using $z$-$(\delta p/p)$ as the longitudinal coordinates. ($c\tau$-$\delta E$ units could also be used.) The first term on the right is the potentially damping term, and longitudinal cooling occurs if the derivative $\partial(dE/ds)/\partial E > 0$. The second term is a heating term due to energy straggling. The energy loss is given by the Bethe-Bloch equation, which we approximate by:

$$\frac{dE}{ds} = 4\pi N_A r_e^2 m_e c^2 \rho \frac{Z}{A}\left[\frac{1}{\beta^2}\ln\left(\frac{2m_e c^2 \gamma^2 \beta^2}{I(Z)}\right) - 1 - \frac{\delta}{2\beta^2}\right] \quad , \tag{4}$$

where $E_\mu = \gamma m_\mu c^2$ is the muon energy, $P_\mu = \beta\gamma m_\mu c^2$ is the muon momentum, $N_A$ is Avogadro's number $Z$, $A$ and $\rho$ are the material atomic charge and number and density, $I(Z)$ is the material ionization energy, and $\delta$ is the density effect factor which is approximated by 0 in the following discussion.

The $(\delta p/p)^2$ scattering term is estimated using the formula for energy straggling:

$$\frac{d(\frac{\delta p}{p})^2_{rms}}{ds} \cong \frac{1}{\beta^2 c^2 p^2}\frac{d(\Delta E_{rms}^2)}{ds} = \frac{1}{\beta^2 c^2 p^2}4\pi(r_e m_e c^2)^2 m_e c^2 n_e \gamma^2\left(1 - \frac{\beta^2}{2}\right)$$

where $n_e \cong \rho N_A Z/A$ is the electron density. This estimate may not be as accurate as desired. It excludes transverse/longitudinal mixing effects.

$g_L$ is the longitudinal partition number, which is approximately given by:

$$g_{L,0} = -\frac{2}{\gamma^2} + 2\frac{(1 - \frac{\beta^2}{\gamma^2})}{\ln\left(\frac{2m_e c^2 \gamma^2 \beta^2}{I(Z)}\right) - \beta^2} \tag{5}$$

This factor must be >0 for cooling, but is in fact negative for $P_\mu <$ ~ 350 MeV/c, and only weakly positive for higher energies (see fig. 2); ionization cooling does not directly provide effective longitudinal cooling.

However, the longitudinal cooling rate can be enhanced by placing the absorbers where transverse position depends upon energy (nonzero dispersion) and where the absorber density or thickness also depends upon energy, such as in a wedge absorber. This makes the beam particle path length through absorber material dependent on energy. (See figure 2.) In that case the cooling derivative can be rewritten as:

$$\frac{\partial \left(\frac{dE}{ds}\right)}{\partial E} \Rightarrow \frac{\partial \left(\frac{dE}{ds}\right)}{\partial E}\bigg|_0 + \frac{1}{\beta^2 E}\frac{dE}{ds}\frac{\eta \rho'}{\rho_0}, \quad (6)$$

where $\rho'/\rho_0$ is the change in density with respect to transverse position, $\rho_0$ is the reference density associated with $dE/ds$, and $\eta$ is the dispersion ($\eta = dx/d(\Delta p/p)$). Increasing the longitudinal cooling rate in this manner decreases the associated transverse cooling by the same amount. The transverse cooling term is changed to:

$$\frac{d\varepsilon_x}{ds} = -\frac{1}{\beta^2 E}\frac{dE}{ds}\left(1 - \frac{\eta \rho'}{\rho_0}\right)\varepsilon_x. \quad (7)$$

Thus, with wedge cooling the longitudinal and dispersion-coupled transverse partition numbers are modified:

$$g_L \Rightarrow g_{L,0} + \frac{\eta \rho'}{\rho_0}; g_x \Rightarrow 1 - \frac{\eta \rho'}{\rho_0}, \quad (8)$$

where the x-coordinate is the transverse dispersion/wedge dimension coupling to the longitudinal motion. More generally, coupling of transverse and longitudinal damping mixes the cooling rates under the constraint that the sum of the cooling rates (damping partition numbers) $\Sigma_g$ is constant, with a momentum dependence:

$$\Sigma_g = g_x + g_y + g_L = 2 + g_{L,0} \cong 2\beta^2 + 2\frac{(1 - \frac{\beta^2}{\gamma^2})}{\ln\left(\frac{2m_e c^2 \gamma^2 \beta^2}{I(Z)}\right) - \beta^2} \quad (9)$$

This momentum dependence is displayed in fig. 2. $\Sigma_g$ is approximately 2 for $P_\mu > 200$ MeV/c, but drops to small values at low momentum. Consequently, most of our cooling channels are designed for $P_\mu \cong 200$—300 MeV/c.

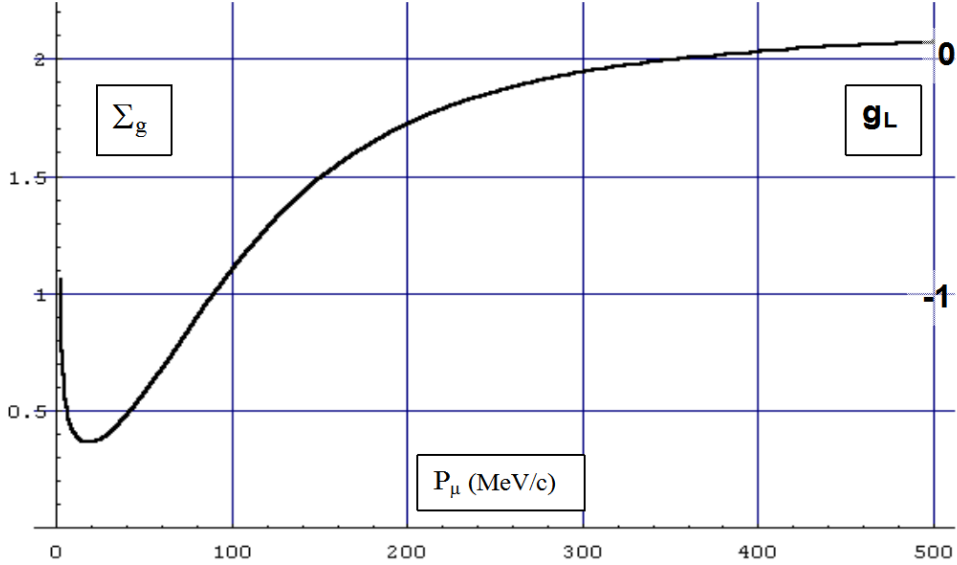

**FIGURE 2** Longitudinal partition number $g_L$ as a function of muon momentum $P_\mu$. The sum of partition numbers $\Sigma_g = 2 + g_L$ is also shown.

The partition number change given by a wedge has a very simple evaluation from the graphical representation presented in Fig.3: $\delta g_L = \eta \rho'/\rho_0 = \eta/w$, where $w$ is the distance from the beam center orbit ($\delta=0$) to the apex of the wedge (see fig.3). This assumes the beam size is less than $w$.

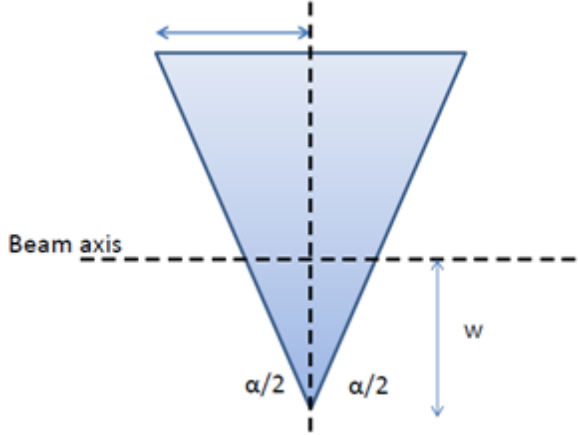

**FIGURE 3.** Wedge geometry for emittance exchange. The beam passes from left to right with its center along the axis ($\delta$=0), with a dispersion $\eta$ at the wedge. The wedge opening angle is $\alpha$ and the distance from beam centroid to apex is $w$. With wedge absorbers $\delta g_L = \eta \rho'/\rho_0 = \eta/w$.

The cooling equations (1) and (3) have characteristic exponential solutions:

$$\varepsilon_i(s) = (\varepsilon_{i,0} - \varepsilon_{i,eq})e^{-s\frac{g_i dP_\mu/ds}{P_\mu}} + \varepsilon_{i,eq} \tag{10}$$

Where $i = x, y$ or $L$, for the appropriate dimension, $\varepsilon_{i,0}$ is the initial emittance, and $\varepsilon_{i,eq}$ is the equilibrium emittance found from balancing the heating and cooling terms. For transverse and longitudinal emittances the equilibrium emittances are:

$$\varepsilon_{t,eq} \cong \frac{\beta_\perp E_s^2}{2 g_t \beta m_\mu c^2 L_R \frac{dE}{ds}}$$ for transverse motion and $$\varepsilon_{L,eq} \cong \frac{\beta_L m_e c^2 \gamma^2 \left(1 - \frac{\beta^2}{2}\right)}{2 g_L \beta m_\mu c^2 \left[\frac{\ln[\frac{2 m_e c^2 \gamma^2 \beta^2}{I(Z)}]}{\beta^2} - 1\right]}$$ for longitudinal motion. Another critical component of the cooling solutions is the cooling length given by:

$$L_{cool,i} = \left[\frac{g_i}{P_\mu}\frac{dP_\mu}{ds}\right]^{-1} = \left[\frac{g_i}{\beta^2 E_\mu}\frac{dE_\mu}{ds}\right]^{-1} \tag{11}$$

where the energy loss is averaged over the full transport length. The cooling length must be much less than the decay length (660 $\beta\gamma$ m); preferably < ~100m.

Cooling to small emittances requires small $\beta_\perp$ (strong focusing) at the absorbers, and the large ($L_R dE/ds$) found in low-Z materials. Cooling to small longitudinal emittance requires small $\beta_L$ and modestly relativistic $\gamma$ ($\gamma = 2$—3). Material properties are summarized in Table 1.

**Table 1: Material Properties for Ionization Cooling**

| Material | Symbol | Z, A | Density | dE/ds (min.) | $L_R$ | $L_R$ dE/ds | $\sigma_0 \cdot \beta\gamma^{1/2}$ | $g_x \beta \varepsilon_N / \beta_\perp$ |
|---|---|---|---|---|---|---|---|---|
| | | | gm/cm$^3$ | MeV/cm | Cm | MeV | | mm-mrad/cm |
| Hydrogen | H$_2$ | 1, 1 | 0.071 | 0.292 | 865 | 252.6 | 0.061 | **37** |
| Lithium | Li | 3, 7 | 0.534 | 0.848 | 155 | 130.8 | 0.084 | **71** |
| Lith. H | LiH | 3+1, 7+1 | 0.82 | 1.60 | 95 | 152 | 0.079 | **60** |
| Beryllium | Be | 4, 9 | 1.848 | 2.98 | 35.3 | 105.2 | 0.094 | **88** |
| Carbon | C | 6, 12 | 2.265 | 4.032 | 18.8 | 75.8 | 0.110 | 122 |
| Aluminum | Al | 13, 27 | 2.70 | 4.37 | 8.9 | 38.9 | 0.154 | 238 |
| Copper | Cu | 29, 63.5 | 8.96 | 12.90 | 1.43 | 18.45 | 0.224 | 503 |
| Tungsten | W | 74, 184 | 19.3 | 22.1 | 0.35 | 7.73 | 0.346 | 1200 |

## Longitudinal motion parameters

Longitudinal cooling depends on the parameter $\beta_L$, which must be properly defined within the present coordinates. Longitudinal motion within rf fields is often more easily expressed in ($c\tau$-$\delta E$), and longitudinal cooling and bunching equations can be developed in those coordinates. For closer resemblance to transverse motion, $\delta z$-$\delta p/p$ coordinates are used here. Since $\delta z \sim \beta c\tau$ and $\delta p/p = \delta E/(\beta^2 E)$, errors by factors of $\beta$ are easily obtained; any correct corrections to the following algebra are welcome.

Longitudinal motion is controlled by rf, where rf cavities are used to recover energy lost in the absorbers and to bunch the beam. To maintain constant energy, the energy lost in the absorber is recovered by rf gradient:

$$\frac{dE}{ds} = eV'\sin(\phi_s)$$

The rf also provides bunching, following the linearized equations of longitudinal motion:

$$\frac{dz}{ds} = -\alpha_p \frac{\delta p}{p} \quad \text{and} \quad \frac{d(\frac{\delta p}{p})}{ds} = \frac{2\pi}{\lambda_{rf}} \frac{eV'\cos(\phi_s)}{\beta^2 P_\mu} z.$$

From these equations we can find the longitudinal betatron function:

$$\beta_L = \sqrt{\frac{\lambda_{rf}\beta^3\gamma m_\mu c^2 \alpha_p}{2\pi eV'\cos(\phi_s)}}$$

where $\alpha_p = 1/\gamma^2$ for a linac and $|1/\gamma^2 - 1/\gamma_t^2|$ for a synchrotron and more generally includes the relative path length dependence on $\delta p/p$ in a "$1/\gamma_t^2$" term. (Sign conventions and phases are chosen to obtain acceleration and bunching at $\phi_s$.) At typical linac parameters ($P_\mu$ = 200 MeV/c, $V'\cos(\phi_s)$ = 10 MV/m, $\lambda_{rf}$ =0.923m), $\beta_L$= 0.708m.

This evaluation assumes linear motion, which implies short bunch length. Initial cooling systems fill the rf bucket, reducing the accuracy of the linearized results.

## Modifications Developed through the Muon Collaboration Cooling Studies

The muon collaboration has explored and developed the initial cooling concepts with simulation and analytical studies, and is developing multiple step cooling scenarios using muon cooling. These studies have advanced the understanding and optimization of the cooling process, and more accurately identified optimal cooling channel features.

At ~200 MeV/c the preferred focusing magnets are solenoids, which focus both x and y and also couple x and y motion. Solenoidal focusing lattices have been developed with relatively small $\beta_t$ at the absorbers and with dispersion at the absorbers that can be combined with wedges to obtain 6-D cooling.[5] Typically, x and y motion is so tightly coupled that they cannot be separated, even though the wedge/dispersion is predominantly in one plane. Thus the partition numbers are adequately approximated by:

$$g_L \rightarrow g_{L,0} + \delta g_L \text{ and } g_x = g_y = 1 - \delta g_L/2$$

Within solenoidal focusing, particles in the beams have angular momentum. To some extent this complication can be ignored if there are periodic field flips (which reverse the sign of the angular momentum). The intrinsic angular momentum is damped by the ionization cooling absorbers.[4]

With moderately relativistic motion and strong solenoidal motion, the longitudinal motion is coupled to the transverse amplitudes. An initial distribution that is generated without correlations is mismatched into a cooling channel and that mismatch is seen as particle losses. Attempts to generate the appropriate correlations in simulated initial distributions have been only partially successful; an accurate analytical representation of optimal correlation is not yet generally known. With ionization damping plus multi-cell transport, an initially uncorrelated beam damps toward a properly correlated beam. Also a beam generated by simulation of the capture and phase energy rotation and initial cooling of neutrino factory/ muon collider systems naturally develops correlations matching those needed by the following cooling channels, usually more accurately than generated by analytical approximations.

Also, while beam optics matching into cooling sections is desirable, matching need not be as precise as initially expected. The ionization cooling process damps betatron mismatches.

*Heating by coupling of transverse and longitudinal*

The equilibrium emittances shown above assume transverse heating is dominated by multiple scattering and longitudinal heating is dominated by straggling, with the two effects decoupled. While that approximation is valid for "most" of the parameter spaces explored, scattering and energy loss are coupled by dispersion at the absorber. Kim and Wang [4] have developed equations for this effect, given by:

$$\frac{1}{\beta\gamma}\frac{d\varepsilon_{\perp,N}}{ds} \to \frac{\beta_{\perp}}{2}\frac{d\langle\theta^2\rangle}{ds} + (\gamma_{\perp}\eta^2 + 2\alpha_{\perp}\eta\eta' + \beta_{\perp}\eta'^2)\frac{d\langle\delta_p^2\rangle}{2\,ds}$$

$$\frac{1}{\beta\gamma}\frac{d\varepsilon_{L,N}}{ds} \to \frac{\beta_L}{2}\frac{d\langle\delta_p^2\rangle}{ds} + (\gamma_L\eta^2 + 2\alpha_L\eta\eta' + \beta_L\eta'^2)\frac{d\langle\theta^2\rangle}{2\,ds}$$

where $\beta_i$, $\alpha_i$, $\gamma_i$ are Courant-Snyder transverse ($\perp$) and longitudinal (L) betatron functions, $\delta_p = \delta p/p$ and $\eta$ is the dispersion. If the wedge-absorber is at an optical waist ($\eta' = 0$, $\alpha_i = 0$) then the factor in the last term on the right is $\eta^2/\beta_i$. For this coupling to be relatively small, it is necessary that $\eta/\beta_i$ be small. ($\eta < \beta_{\perp}$ and $\eta < \beta_L/3$.) If $\eta'$ is non-zero, $\eta'$ should also be small ($\eta' < \sim 0.3$).

# EVALUATION OF COOLING EQUATIONS AND COMPARISONS WITH SIMULATIONS

The exponential solution of emittance as a function of time (or distance traveled) has been found and displayed in innumerable cooling simulation and studies, although a direct comparison between simulated and analytically expected results is not generally presented. The simulations have generally followed analytic expectations; however, more direct and quantitative comparisons should be more generally undertaken for several reasons:

1. Ionization cooling performance must be close to expected behavior for the cooling to be useful for a future collider; there is very little safety margin within the goal of a high luminosity collider. Channels significantly below desired performance are inadequate and should be identified.

2. Deviations from expected performance are evidence of deficiencies in the cooling channel and/or the cooling model. Accurate comparisons are needed to identify the sources and possible mitigation of these effects.

An important feature that must be differentiated from cooling is beam loss by scraping, either by collimation or dynamic aperture losses. These losses remove large amplitude particles and may appear to "cool" the surviving beam, which has a smaller rms emittance. While useful for forming compact beams for a collider, scraping is not cooling and should be properly distinguished.

In initial cooling sections, and in transitions between rf sections, the beam may longitudinally fill the "rf bucket". Small perturbations, as well as straggling and scattering, would then cause beam loss, without emittance increase and perhaps rms emittance decrease. The combination of beam loss with cooling can also magnify the apparent amount of cooling; detailed analysis can identify this effect.

Some yardsticks have been developed for evaluating cooling channels. One is the quality factor $Q$, defined locally by:

$$Q = \frac{\frac{1}{(\varepsilon_x\varepsilon_y\varepsilon_z)}\frac{d(\varepsilon_x\varepsilon_y\varepsilon_z)}{ds}}{\frac{1}{N}\frac{dN}{ds}}$$

which is a useful guideline for cooling rate evaluation, and some cooling channels have large $Q$. However, since collider luminosity $L$ is proportional to $N^2/(\beta_{\perp}(\varepsilon_x\varepsilon_y)^{1/2})$, $Q > \sim 6$ is needed to break even. Good cooling channels must have very large $Q$.

$dN/ds$ is proportional to muon decay and depends also on beam loss due to large amplitude scatters and dynamic aperture losses. $dN/ds$ is also dependent on the initial beam distribution, and can be exaggerated by poorly matched beam or lessened by an unrealistically good match. Separation of the causes of losses is not always easy.

Another criterion with some validity is $g_{eff}$, an effective total cooling rate generalized from the partition numbers. For a cooling channel segment of length L, $g_{eff}$ is given by :

$$g_{eff} = \frac{\ln\left[\frac{(\varepsilon_x \varepsilon_y \varepsilon_z)_{start}}{(\varepsilon_x \varepsilon_y \varepsilon_z)_{end}}\right]}{\frac{L}{P_\mu}\frac{dP_\mu}{ds}}.$$

At any particular part of a cooling channel a local value of the cooling rate can be defined by the differential:

$$g_{eff,L}(s) = -\frac{\frac{1}{(\varepsilon_x \varepsilon_y \varepsilon_z)}\frac{d(\varepsilon_x \varepsilon_y \varepsilon_z)}{ds}}{\frac{1}{P_\mu}\frac{dP_\mu}{ds}}$$

From the above analysis, $g_{eff}$ must be less than $2 + g_{L,0}$ which is ~1.7 at $P_\mu = 200$ MeV/c. $g_{eff}$ is reduced by any heating effects. An efficient cooling channel should have $g_{eff} > 1.0$. A cooling channel with $g_{eff} < 0.5$ is significantly less efficient than desired. When the initial emittance is close to the equilibrium emittance, $g_{eff}$ becomes small.

## Evaluation of typical cooling channel segments

In the follow subsections we discuss particular segments of cooling channels and evaluate them using the above criteria.

### *Front End Cooling Segment*

The baseline Front End of the neutrino factory has a transverse cooling segment consisting of LiH absorbers with 325 MHz rf cavites and solenoidal focusing.

**TABLE 2.** Parameters of Front End Cooling.

| Parameter | Symbol | Value |
| --- | --- | --- |
| Beam momentum | $P_\mu$ | 245 MeV/c |
| Length of Cooling section | L | 50 m |
| Absorber / Cell Length | $L_{abs}$, $L_C$ | 3cm LiH, 0.75m |
| Maximum magnetic field | $B_{max}$ | 2.8T |
| RF (frequency, gradient, occupancy) | $f_{RF}$, V', f | 325 MHz, 25 MV/m, 0.67 |
| CS values at absorber | $\beta_\perp$, $\eta$ | 0.8m, 0 |
| Wedge parameters | $\alpha$, w | No wedge |
| Partition numbers | $g_x$, $g_y$, $g_L$ | 1, 1, -0.18 |
| Effective g, cooling length | $g_{eff}$, $L_{cool}$ | 0.9, 40/$g_i$ m |
| Equilibrium emittance | $\varepsilon_{t,0}$, $\varepsilon_{L,0}$ | 0.0051m, 0.0005/$g_L$ m |
| Initial, final transverse emittance | $\varepsilon_{t,1}$, $\varepsilon t_{t,f}$ | 0.0142, 0.0078m |

Transverse emittance cooling in both x and y by a factor of 2 is obtained in a 50m channel, in close agreement with rms equation expectation.

There is no wedge/dispersion coupling here, so longitudinal cooling is not expected. Since the dE/ds is anti-damping with $g_{L,0} = -0.18$, an ~15% increase in longitudinal emittance is expected. In simulation the $\varepsilon_L$ remains ~constant emittance. This is due to the fact that rf buckets are ~full at the end of the μ capture region, and particles that are scattered to larger amplitudes are lost from the capture. ~15% of captured muons are lost in this process. This loss is doubled if the μ capture momentum is reduced to 200 MeV/c, where $g_{L,0} = -0.3$. Consideration of this loss effect has led to the increase in capture momentum from ~200 MeV/c in previous versions to the present value of 245 MeV/c.

## RFOFO Cooling Channels from Balbekov

Balbekov has developed a number of cooling channels based on solenoidal focusing, with tilted coils that introduce a small dispersion useable for emittance exchange.[6,7] Fig. 4 displays a particular example, and Table 3 displays parameters of that example and another and summarizes simulation/cooling performance calculations.

In the first example, The central beam momentum is at 240 MeV/c, where $g_{L,0}$ =-0.18. While the wedge angle at 60° seems large, because of the small dispersion the increase in partition number $\delta g_L$ is only 0.33, so the net cooling partition is $g_L$=0.15. The system cools quite well transversely and the overall $g_{eff}$ is ~ 1.25 for a 60m long cooling system, assuming injected emittances of $\varepsilon_t$=0.0025 and $\varepsilon_L$=0.003.

The second example is for initial cooling of a relative large emittance beam and has larger $\beta_t$ and $\eta$ (40cm and 8.5cm). A 100m segment cools $\varepsilon_t$ from ~0.02 to 0.004 m and $\varepsilon_+$ from 0.02 to 0.01m.

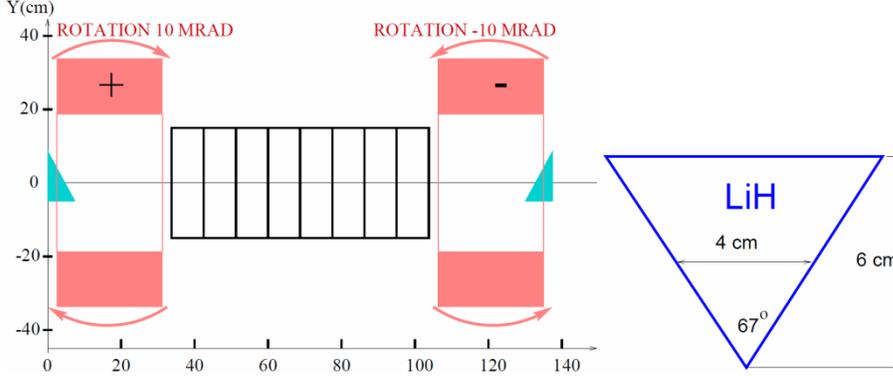

**FIGURE 4.** RFOFO Cooling channel with expanded view of wedge absorber.

**TABLE 3.** Parameters of Balbekov RFOFO coolers

| Parameter | Symbol | Example 1 | Example 2 |
|---|---|---|---|
| Beam momentum | $P_\mu$ | 240 MeV/c | 200 MeV/c |
| Length of Cooling section | $L$ | 60 m | 100m |
| Absorber / Cell Length | $L_{abs}, L_C$ | 3cm LiH, 0.75m | 21.8cm $LH_2$, 1.0 |
| RF (frequency, gradient, occupancy) | $f_{RF}, V', f$ | 800 MHz, 22 MV/m, 0.67 | 325 MHz, 25 MV/m, 0.75 |
| Maximum Magnetic field | $B_{max}$ | 3.5T | 3.5T |
| CS values at absorber | $\beta_\perp, \eta$ | 0.082m, 1cm | 0.4m, 8.5cm |
| Wedge parameters | $\alpha, w$ | 67°, 3cm | 57.1°, 22cm |
| Partition numbers | $g_x, g_y, g_L$ | 0.83, 0.83, 1, 0.15 | 0.78, 0.78, 0.16 |
| Effective g, cooling length | $g_{eff}, L_{cool}$ | 1.34, 40/$g_i$ m | 1.0, 23/$g_i$ m |
| Equilibrium emittance | $\varepsilon_{t,0}, \varepsilon_{L,0}$ | 0.00051m, 0.0016 m | 0.0018, 0.0015 m |
| Initial, final transverse emittance | $\varepsilon_{t,1}, \varepsilon t_f$ | 0.0025, 0.0012m | 0.02, 0.004 |
| Initial, final longitudinal emittance | $E_{L,1}, \varepsilon_{Ltf}$ | 0.003, 0.00275m | 0.02, 0.01 |

## RFOFO Cooling Channels from Stratakis and Palmer

Palmer and Stratakis[8] have developed a sequence of 16 RFOFO cooling channels, based on Balbekov's design and on previous "Guggenheim" designs[9], that reduce transverse emittances by a factor of 10 and longitudinal emittances by a factor of 5. The total length of the system is 560m. The individual sections vary in length from 17 to 83m (typically ~40m) and are composed of individual cells of alternating solenoids, tilted to introduce some dispersion, with wedge absorbers and rf reacceleration. The cell lengths vary from 2.75m to 0.806m and the focusing field increases from B=2.7T to 13T while betatron functions decrease: $\beta_t$ =0.40 to 0.04m and $\eta$=6.6 → 0.6cm. In simulations the transverse emittance decreases from 3.65 to 0.32 mm while the longitudinal emittance decreases from 9.68 to 1.6 mm. (The last increment in longitudinal cooling from 2 to 1.6mm is due to beam loss rather than cooling.) Table 4 presents parameters of two of these RFOFO cooling channels.

With the relatively small dispersion, relatively large wedge angles are needed to include longitudinal cooling, and α =100--120° is used. One limitation of the system is that the effective cooling number is only $g_{eff}$= 0.4. This is largely a result of the design choice to operate in each section with the transverse emittance fairly close to the equilibrium emittance. In a typical segment the emittance is only 20 to 40% larger than the equilibrium value. In segments where the emittance is 60-100% larger than equilibrium, $g_{eff}$ increases to ~0.8.

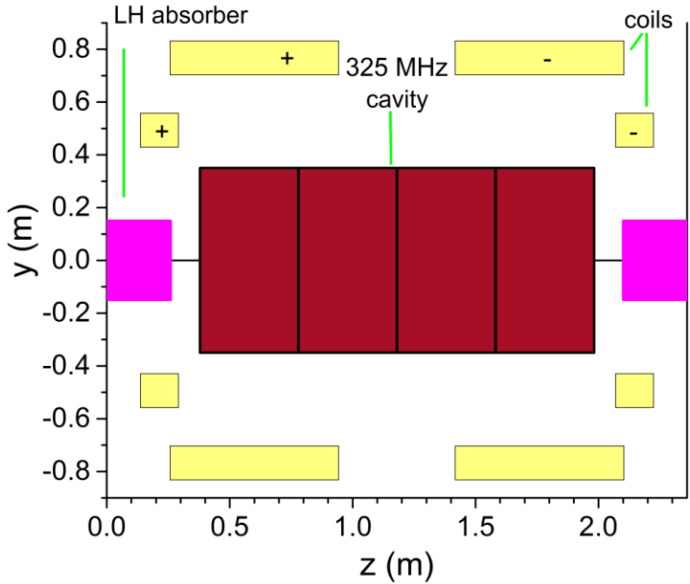

**FIGURE 5.** An RFOFO Cooling channel cell for the Palmer/Stratakis scenario.

**TABLE 4.** Parameters of Palmer Stratakis RFOFO coolers

| Parameter | Symbol | Section 2 | Section 10 |
|---|---|---|---|
| Beam momentum | $P_\mu$ | 204 MeV/c | 204 MeV/c |
| Length of Cooling section | $L$ | 33 m | 83m |
| Absorber / Cell Length | $L_{abs}$, $L_C$ | 26cm $H_2$, 2.36m | 13cm $H_2$,0.81m |
| RF (frequency, gradient, occupancy) | $f_{RF}$, $V'$, $f$ | 325 MHz, 17.5 MV/m, 0.67 | 650 MHz, 27 MV/m, 0.67 |
| Magnetic field | $B_{max}$ | 3.1T | 9.5T |
| CS values at absorber | $\beta_\perp$, $\eta$ | 36cm, 6.8cm | 8.7cm, 2.3cm |
| Wedge parameters | $\alpha$, $w$ | 100°, 11cm | 110°, 3.5cm |
| Partition numbers | $g_x$, $g_y$, $g_L$ | 0.74, 0.74, 0.24 | 0.77, 0.77, 0.18 |
| Effective g, cooling length | $g_{eff}$, $L_{cool}$ | 0.45, 51/$g_i$ m | 0.41, 36/$g_i$ m |
| Equilibrium emittance | $\varepsilon_{t,0}$, $\varepsilon_{L,0}$ | 0.0027m, 0.0022 m | 0.0006, 0.0019 |
| Initial, final transverse emittance | $\varepsilon_{t,1}$, $\varepsilon t_f$ | 0.00342, 0.00317m | 0.00096, 0.00067 |
| Initial, final longitudinal emittance | $E_{L,1}$, $\varepsilon_{Ltf}$ | 0.0078, 0.0067m | 0.0037, 0.0030 |

*Helical Cooling Channels from Derbenev and Yonehara et al.*

Derbenev and Johnson[10] proposed the use of a helical geometry for a cooling channel with constant solenoid, and helical dipole and quadrupole fields. The fields place particles on a strong-focusing helical orbit which has an energy-dependent path length. Key parameters of the helical system include the period of the helix $\lambda$, the solenoidal field $B$, the helical field $b$, and the radius of the helical orbit $a$, which is momentum-dependent, with that dependence given by the equation:

$$p(a) = \frac{\sqrt{1+(ka)^2}}{k}\left[B - \frac{1+(ka)^2}{ka}b\right]$$, where $k = 2\pi/\lambda$, and $p$ is in units of T-m. The pitch $\kappa$ of the helix is: $\kappa = 2\pi a/\lambda = ka$. A variety of helical channel parameters may be obtained.

That energy dependent path length is obtained from the dispersion D on the helical orbit, given by solving:

$$\frac{a}{D} = \frac{2(ka)^2 - 1}{1+(ka)^2} + \frac{(1-(ka)^2)}{\sqrt{1+(ka)^2}}\frac{B}{kp} - \frac{(1+(ka)^2)^{\frac{3}{2}}}{k^2 p}\frac{\partial b}{\partial a}$$

From that variation in the path length with momentum, an enhancement of longitudinal cooling is obtained, given by:

$$\delta g_L = \frac{\kappa^2}{1+\kappa^2}\frac{D}{a}$$

The path length also changes the longitudinal motion following:

$$\alpha_p = \frac{\kappa^2}{1+\kappa^2}\frac{D}{a} - \frac{1}{\gamma^2}$$

A variety of helical channel parameters may be obtained. Derbenev presents a number of cases beginning with setting $\kappa=1$, which implies $a = \lambda/2\pi$. A further requirement of equal transverse and longitudinal cooling at 200 MeV/c then requires $\delta g_L \cong 0.8$ or $D \cong 1.6a$.

Yonehara et al[11] have constructed a 6-step cooling scenario based on this concept with the focusing strength increasing from segment to segment, correlated with smaller $\beta_t$. Table 5 displays the features of each section of the cooling channel with some results on simulation of that cooling sequence. The effective cooling factor $g_{eff}$ is also shown.

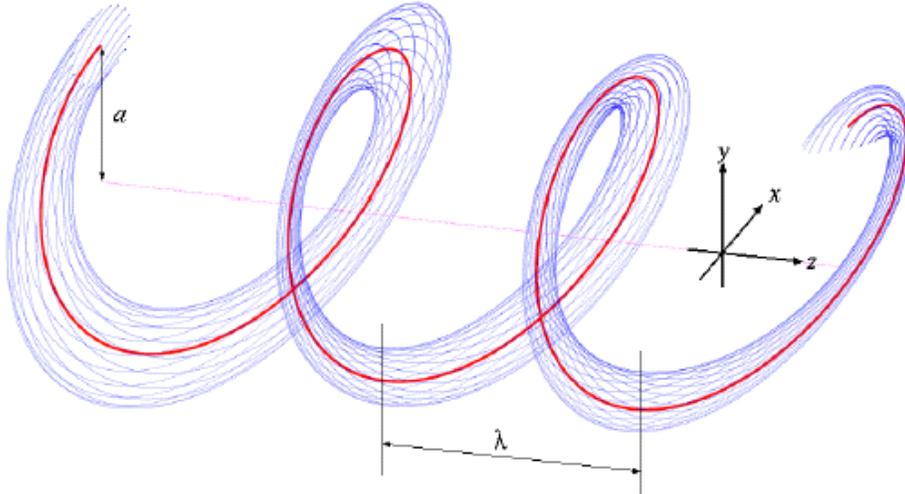

**FIGURE 6** Particle orbits through a helical cooling channel. The red line is the central beam orbit and the blue lines are particle orbits within the beam.

There are two anomalies in the table. Section 2 shows a $g_{eff}$ of 2.4, when the largest $g_{eff}$ possible from cooling is ~1.7. This is a short "cooling" section with relatively large beam loss from scraping; the emittance decrease is largely from beam loss rather than true cooling. Section 4 has a relatively small $g_{eff}$ (~0.4). That section was poorly matched from the previous section (partly from the 325→650 MHz transition.). A better match between the sections has since been obtained and the cooling section length is reduced from ~90 m to ~50m with the same degree of cooling; $g_{eff}$ is then increased to ~0.8.

As with other scenarios the last cooling section is relatively inefficient ($g_{eff} = 0.54$) as the emittances approach equilibrium values. All of the cooling scenarios discussed in this section can cool the transverse emittance to ~0.3mm, and are limited by focusing strength to that approximate value. (The maximum magnetic fields used are ~14T.) This is the emittance level required by a $\mu^+$-$\mu^-$ Higgs Collider.

**TABLE 5.** Parameters of a helical cooling channel.

| unit | $L$ m | $b_z$ T | $\lambda$ m | $\beta_{T,eff}$ cm | $D$ m | $\nu$ GHz | $\varepsilon_{T,eq}$ mm | $\varepsilon_T$ mm | $\varepsilon_L$ mm | $\varepsilon_T^2 \varepsilon_L$ mm$^3$ | $g_{eff}$ | $\varepsilon$ Transmission |
|---|---|---|---|---|---|---|---|---|---|---|---|---|
| 0 | 0 | | | | | | | 20.4 | 42.8 | 17800 | | |
| 1 | 40 | -4.2 | 1.0 | 14 | 0.3 | 0.325 | 1.0 | 5.97 | 19.7 | 702 | **1.65** | 0.92 |
| 2 | 9 | -4.8 | 0.9 | 12 | 0.27 | 0.325 | 0.9 | 4.01 | 15.0 | 241 | **2.43** | 0.86 |
| 3 | 80 | -5.2 | 0.8 | 10 | 0.24 | 0.325 | 0.8 | 1.02 | 4.8 | 4.99 | **0.99** | 0.73 |
| 4 | 90 | -8.5 | 0.5 | 7 | 0.15 | 0.65 | 0.45 | 0.58 | 2.1 | 0.706 | **0.44** | 0.66 |
| 5 | 24 | -9.8 | 0.4 | 5.5 | 0.12 | 0.65 | 0.3 | 0.42 | 1.3 | 0.229 | **0.96** | 0.64 |
| 6 | 30 | -14 | 0.3 | 4.0 | 0.09 | 0.65 | 0.24 | 0.32 | 1.0 | 0.104 | **0.54** | 0.62 |

Cooling to smaller emittances requires more extreme focusing and phase-space manipulation than the presently discussed 6-D cooling systems with solenoidal focusing. (Concepts for these include Li-lens based cooling, low-energy with emittance exchange, reverse emittance exchange at extreme parameters.) Those systems are outside the scope of the present discussion. However, these cooling values are adequate by themselves for a $\mu^+$-$\mu^-$ Higgs Collider.[12]

## ACKNOWLEDGMENTS

We thank D. Stratakis, M. Palmer, R. Palmer, K. Yonehara, V. Balbekov and C. Yoshikawa for helpful discussions. This research was supported by the US DOE under contract DE-AC02-07CH11359.